\documentclass[twocolumn,twoside,showkeys,showpacs,preprintnumbers,superscriptaddress,amsmath,amssymb,prc]{revtex4-2}

\usepackage{graphicx}
\usepackage{dcolumn}
\usepackage{bm}
\usepackage[version=4]{mhchem}
\usepackage{subcaption}
\usepackage[colorlinks=true,linkcolor=black,citecolor=blue,urlcolor=blue,]{hyperref}
\usepackage{appendix}
\usepackage{balance}
\usepackage{ulem}
\usepackage[justification=raggedright,singlelinecheck=false]{caption}

\DeclareUnicodeCharacter{0394}{$\Delta$}
\DeclareUnicodeCharacter{03B3}{$\gamma$}

\def\thu{Department of Physics, Tsinghua University, Beijing 100084, China}
\def\hepcent{Center of High Energy Physics, Tsinghua University, Beijing 100084, China}

\begin{document}

\title{Imaging Freeze-out Sources and Extracting Strong Interaction Parameters in Relativistic Heavy-Ion Collisions}

\author{Junhuai Xu} 
\email{xjh22@mails.tsinghua.edu.cn}
\affiliation\thu

\author{Zhi Qin} 
\email{qinz18@mails.tsinghua.edu.cn}
\affiliation\thu

\author{Renjie Zou} 
\affiliation\thu

\author{Dawei Si} 
\affiliation\thu

\author{Sheng Xiao} 
\affiliation\thu

\author{Baiting Tian} 
\affiliation\thu

\author{Yijie Wang} 
\email{yj-wang15@tsinghua.org.cn}
\affiliation\thu

\author{Zhigang Xiao} 
\email{xiaozg@tsinghua.edu.cn}
\affiliation\thu
\affiliation\hepcent

\date{\today}

\begin{abstract}
By combining femtoscopic interferometry with an optical deblurring algorithm, we present a novel method to image the source in heavy-ion collisions (HICs) while simultaneously determining the interaction strength between particle pairs. The spatial distribution of the emission source  has been reconstructed for protons ($p$) and antiprotons ($\bar{p}$) from the respective $pp$ and $\bar{p}\bar{p}$ correlation functions in Au+Au collisions at $\sqrt{S_{\rm NN}}=200$ GeV.  Within experimental uncertainties, protons and antiprotons share the same freeze-out distribution showing higher density in the center compared to the widely assumed Gaussian shape. The results evidence  the matter-antimatter symmetry in coordinate space at the moment of freeze-out before the nucleons are fully randomized in the collisions.

\end{abstract}


\maketitle

{\it Introduction -} It is well established that matter and antimatter were created in equal amounts during the first moments after the Big Bang. However, the observable universe consists overwhelmingly of matter, posing a fundamental question in physics. This puzzling phenomenon is related to the profound symmetry breaking in nature and remains a challenge in nuclear and particle physics. On modern accelerators, heavy ion collisions (HICs) create conditions akin to those in the early universe. So after the success of observing the production of antimatter \cite{Abelev2010,Agakishiev2011,Abdulhamid2024}, people can further investigate the spatial evolution and decoupling of the matter and antimatter on the femtoscopic scale.  

Hanbury Brown and Twiss (HBT) intensity interferometry, originally developed to measure the angular radius of Sirius \cite{Brown1956}, has been extensively adapted in nuclear physics \cite{Wang2022,Wang2021,Qiao2024}. It enables the extraction of spatiotemporal characteristics of the emission source in HICs by analyzing particle-pair correlation functions (CFs) at small relative momenta \cite{Wang2024a,Wang2023}. Additionally, the spatial information embedded in the CFs has been applied to investigate the exotic structure and rare decay mode of unstable nuclei \cite{Zhou2024,Revel2018,Laurent2019,Cao2012,Zhou2022}. In a basic picture, the CF is the result of the low-energy scattering between the particle pair emitted from the source. It carries the information of the interaction strength, defined by the scattering length $f_0$ and effective range  $d_0$ between the pair, written as ($f_0$, $d_0$) in short.

Thus, simultaneous inference of the source distribution and the interaction strength becomes feasible for either particle or antiparticle pairs. Based on the  Lednick\'y-Lyuboshitz (LL) model, STAR Collaboration has verified that protons and antiprotons exhibit nearly identical interaction properties, and the extracted source size parameters for  $pp$  and  $\bar{p}\bar{p}$  pairs were found to be similar \cite{Adamczyk2015}. Specifically, within experimental uncertainties, the parameters  ($f_0$,  $d_0$) are equivalent for  $pp$  and  $\bar{p}\bar{p}$  pairs, consistent with CPT symmetry, a fundamental principle in physical laws. The Gaussian source assumption has become widely adopted in correlation function analysis, from which significant results have been obtained in the studies of  particle interactions and source distributions  \cite{Goldhaber1960,Boal1986,Boal1990,Verde2002}.

However, more studies start to notice the non Gaussian feature of the source. By Gaussian parameterization it is assumed that all the participants are totally randomized in space prior to the freeze-out instant, and  the initial state as well as the evolution details of the colliding system are lost. This strong assumption is unfavorable in the context of various recent studies. For two examples, i) high energy HIC has been applied to infer the initial shape of the colliding nuclei \cite{Collaboration:2024zme,Zhao2024}.  ii)  The Ohm's law is suggestively broken because of the strong field and extremely short evolution time \cite{Wang:2021oqq}. In both cases it is of importance to probe the spatial evolution of the colliding system. In recent experiments, a similar deviation from Gaussian distributions has been observed in the freeze-out source distributions of particles, exhibiting long tails and Lévy-stable characteristics \cite{Adare2018,Mukherjee2023,Kovacs2023,Kincses2024,Abdulameer2024,Kincses2025}. This behavior suggests that hadronic evolution could be described by Lévy walks, where the movement patterns are governed by long-tailed random walks.  This concept has already played a significant role in the natural sciences \cite{Tsallis1995,Zaburdaev2015} and is gradually gaining widespread attention in the field of heavy-ion collisions. Thus, it is crucial to obtain the true distribution without pre-assuming a specific form such as the Gaussian.

Despite of the difficulty in solving inverse problem,  the direct imaging of the source has been achieved without introducing the analytic parameterization \cite{Brown1997,Chung2003,Brown2005}. More recently, the Bayessian-based deblurring method, known as Richardson-Lucy (RL) algorithm \cite{Richardson1972,Lucy1974} which was initially developed for optical imaging \cite{Vankawala2015,Liu2025}, has  gained prominence in nuclear physics \cite{Danielewicz2022,Nzabahimana2023,Nzabahimana2023a,Vargas2013,Xu2024}. Particularly, the RL algorithm has been applied in source imaging in HICs, highlighting the method’s effectiveness and robustness \cite{Nzabahimana2023,Tam2025}. In this letter, combining the RL deblurring algorithm and the LL model incorporating the pair interaction, we bring the realisation to solve the inverse problem of source imaging from the experimental CFs. The source distribution and ($f_0$, $d_0$) are determined simultaneously. The similarity of the source spatial distribution of $p$ and $\bar p$ emission has been demonstrated.

{\it Methodology -} Defined as a function of the relative momentum  $\mathbf{q} = \mathbf{p}_1 - \mathbf{p}_2$  in the particle pair rest frame (PRF), the CF is given experimentally by
$C(\mathbf{q}) =\frac{P(\mathbf{p}_1, \mathbf{p}_2)}{P(\mathbf{p}_1) P(\mathbf{p}_2)}$,
where  $P(\mathbf{p}_1)$  and  $P(\mathbf{p}_2)$  are the probabilities of detecting particles with momenta  $\mathbf{p}_1$  and  $\mathbf{p}_2$, respectively, and  $P(\mathbf{p}_1, \mathbf{p}_2)$  is the joint probability of detecting both particles simultaneously. For uncorrelated emissions,  $C = 1$  is expected. In theoretical frameworks, the correlation function (CF) is computed using the Koonin-Pratt (KP) formula \cite{Koonin1977, Pratt1990}\, 

\begin{equation}\label{KP}
C(\mathbf{q}) = \int d^3r |\Psi_{\mathbf{q}}(\mathbf{r}^{\prime})|^2 S(\mathbf{r})
\end{equation}
where  $S$ is the source function (SF) related to the relative distance $\mathbf{r}=\mathbf{r}_1-\mathbf{r}_2$ of the particle-pair in the source rest frame, and  $\Psi_{\mathbf{q}}(\mathbf{r}^{\prime})$  is the wave function calculated from the relative momentum  $\mathbf{q}$  and distance  $\mathbf{r}^{\prime}$  in the PRF. Based on the LL model \cite{Lednicky1981}, the interaction strength between particle pair is characterized by ($f_0$, $d_0$) describing the strong interaction scattering amplitude in S-wave expansion under the effective range approximation.

\begin{figure}[hptb]
    \centering
    \includegraphics[width=0.9\linewidth]{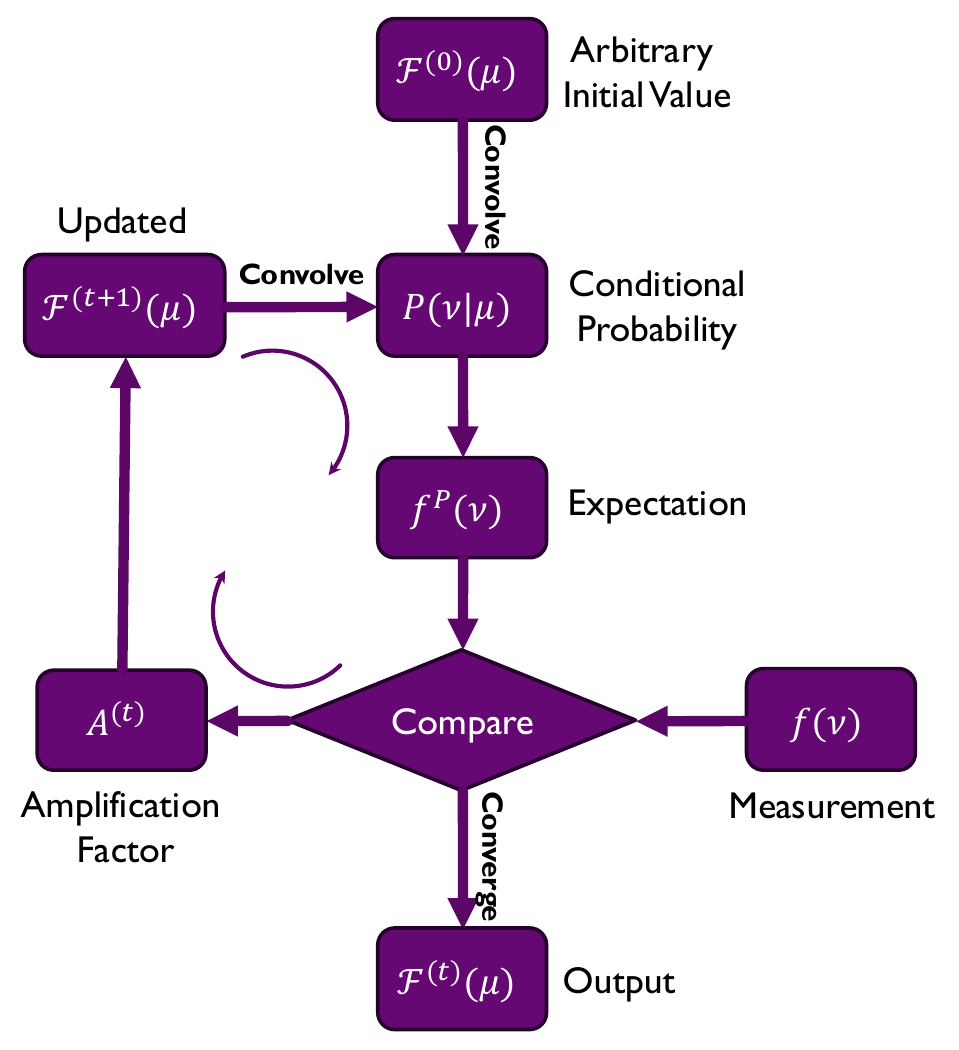}
    \caption{Diagram of RL algorithm process.}
    \label{RLDiagram}
\end{figure}

In optical deblurring, the measured distribution ($f(\nu)$) is related  to the true photon distribution  ($\mathcal{F}(\mu)$)  through a conditional probability function  ($P(\nu|\mu)$), 

\begin{equation}\label{RLdeb}
f(\nu) = \int P(\nu|\mu) \mathcal{F}(\mu) d\mu
\end{equation}
 where  the response function $P(\nu|\mu)$  represents the conditional probability that a photon with the true property  $\mu$  is measured as  $\nu$. Although the true distribution  $\mathcal{F}(\mu)$  is unknown, it can be inferred from the measured distribution  $f(\nu)$  through iterative regression starting from an arbitrary initial guess. This iterative process is illustrated in Fig.~\ref{RLDiagram}. Starting with $t=0$, the arbitrary initial trial  $\mathcal{F}^{(0)}(\mu)$ is convolved  with  $P(\nu|\mu)$  to predict the measurement distribution  $f^P(\nu)$, which  is then compared to the known measurement distribution  $f(\nu)$. Based on the difference between them, an amplification factor  $A^{(t)}$  is calculated at each $t^{\rm th}$ iteration. The factor $A^{(t)}$ is functioning to adjust the values of  $\mathcal{F}^{(t)}(\mu)$  for the next iteration. Through the iterative process, the updated distribution  $\mathcal{F}^{(t+1)}(\mu)$  progressively converges toward the true distribution  $\mathcal{F}(\mu)$. The convergence criteria is invoked if $f^P(\nu)$ becomes sufficiently close to  $f(\nu)$.

From the mathematical similarity between the KP formula (\ref{KP}) and the optical deblurring equation (\ref{RLdeb}), it is naturally suggested that the RL algorithm can be applied iteratively to solve the SF. By evaluating the converged $\chi^2$ value, one can identify the optimal  ($f_0$,  $d_0$)  for the wave function that best describes the CF, thereby achieving both SF imaging and extraction of strong interaction parameters simultaneously. For the convenience  of numerical implementation, the KP formula is  discretized by dividing  $C$  into  $N$  bins and  $S$  into  $M$  bins ($N\ge M$). After discretization, the 
RL algorithm is used to iteratively solve for the distribution of the source function \cite{Nzabahimana2023,Xu2024}. The CF $C$  is expressed by  $S$  through an  $N \times M$  transformation matrix  $K$  as  $C_i = \sum_{j=1}^{M} K_{ij} S_j$. This transformation matrix is calculated by

\begin{equation}\label{KijInt}
K_{ij}= \int_{\text{bin}_j} d\mathbf{Q} |\Psi_{\mathbf{q}_i}(\mathbf{r}')|^2r^2\sin\theta  drd\theta d\varphi.
\end{equation}
where  $\mathbf{Q}$  denotes the total momentum of the particle pair in the source rest frame. Here,  $i$  and  $j$  label the bin indexes of  $C$  and  $S$ , respectively. In one dimensional imaging, the wave function is assumed to be uniform in the polar $\theta$ and azimuthal $\varphi$ directions, leaving only the  $r$-direction to consider. Consequently, the integration over  $\theta$  and  $\varphi$  can be performed using Monte Carlo sampling uniformly within the spherical shell. This approach allows for iterative reconstruction of the source distribution from the discretized experimental measurements of  $C$.

\begin{figure}[hptb]
    \centering
    \includegraphics[width=1.0 \linewidth]{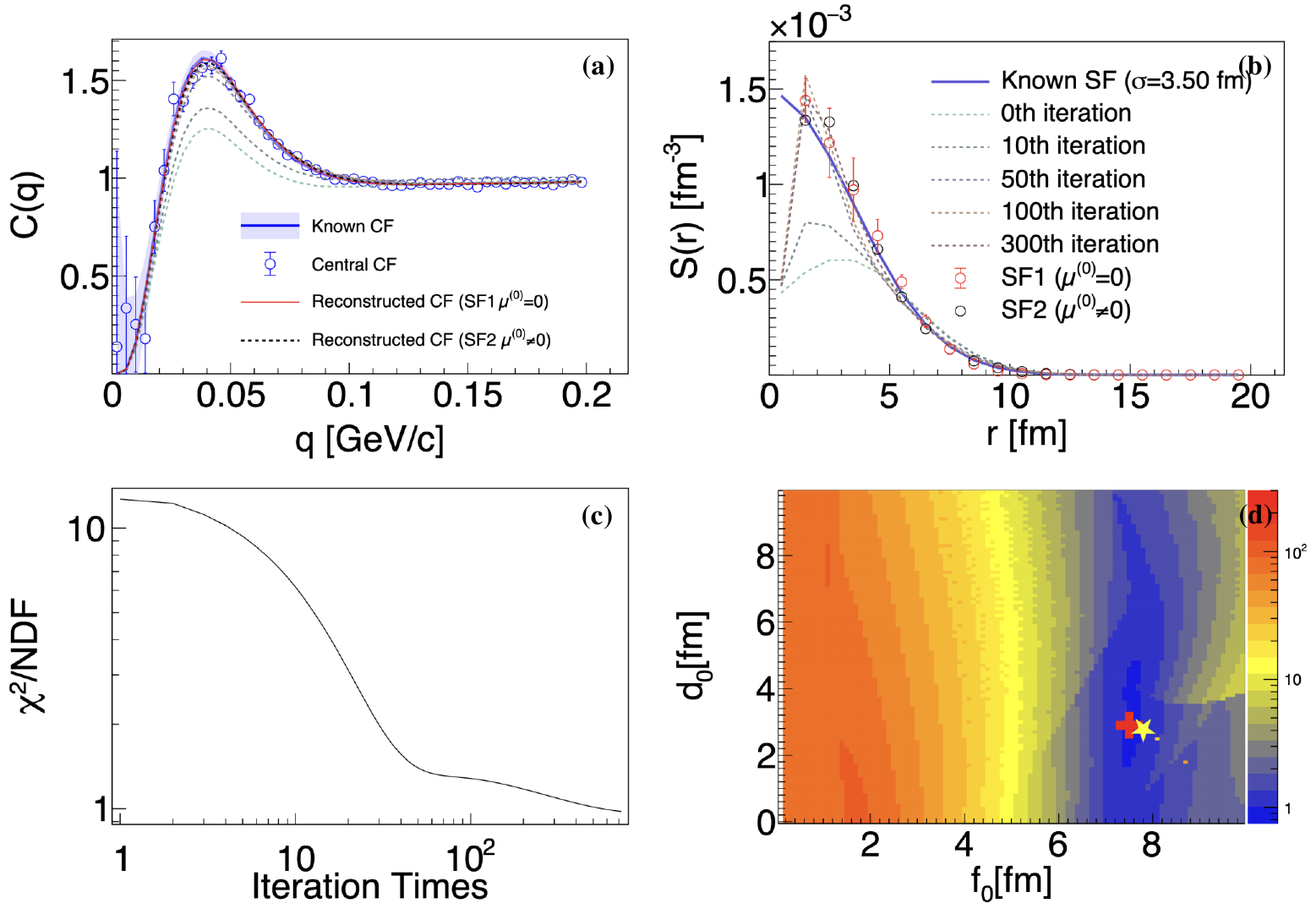}
    \caption{(Color Online) Model test procedure.  (a) The correlations functions. The blue solid curve is the CF calculated using LL model. The symbols are the sampled CF taking into account the uncertainty to mimic the experimental measurement which is selected as the central CF for imaging.  The dashed curves are the output at various iteration steps. The red solid and black dashed thin curves are the final reconstructed CF with two different initial trials of the source function SF1 and SF2. (b) The restored source images. The blue solid curve is the given SF to calculate the CF. The  red and black symbols are the restored SF at two initial trials SF1 and SF2. (c) The $\chi^2/{\rm NDF}$ as a function of iteration times. (d) The minimum $\chi^2/{\rm NDF}$ on the ($f_0$, $d_0$) plane. The yellow star and red cross denote the input value and the final converged solution.}
    \label{ModelTest}
\end{figure}

{\it Model test -}The imaging of the source  and the extraction of interaction parameters ($f_0$,$d_0$) can now be achieved simultaneously using this method. To test the algorithm, we start with calculating the CF from the given source distribution and known $pp$ interactions with $f_0=7.8$\,fm and $d_0=2.77$\,fm. In order to mimic the real measurement, the fluctuation is introduced to each point according to the experimental uncertainty. Fig.~\ref{ModelTest} (a) shows the theoretic CF (thick solid curve) and the fluctuated CF (open circle).  Then the inverse problem is iteratively solved from  a initial trial of the SF, which is taken as a Gaussian distribution  with $\mu=0$ center location and arbitrary deviation. The reconstructed SF ends close to the known one, as marked SF1 in Fig.~\ref{ModelTest} (b). In order to check the robustness of algorithm against the variation of the initial SF, we also test  the second initial trial, where the center of the initial Gaussian is arbitrarily set at  $\mu=3$ fm instead of zero peaking.   With the iteration going on, the  output SF gradually approaches to the known one, as illustrated by the dashed curves in Fig. ~\ref{ModelTest} (b).  The marked SF2 sits on top of SF1, evidencing the robustness of the iterative procedure.   Fig.~\ref{ModelTest} (c) presents the closeness quantity  $\chi^2$ over iteration times in the imaging process of the central CF. Here   $\chi^2$ is defined as $\chi^2 = \sum_i^{N} \left(\frac{C_i^{P}-C_i}{\sigma_i}\right)^2$,
where $C_i^{P}$ represents the value of the $i^{\rm th}$ point. During the procedure, ($f_0$,$d_0$) parameters are optimized by minimum $\chi^2$ and the closeness between the output and the known CF is checked in every iteration till the convergence condition is reached, namely the change of $\chi^2$ between two successive iterations is less than $1 \times 10^{-4}$ , which functions as a threshold  to avoid noise amplification. Fig.~\ref{ModelTest} (d) shows the minimum  $\chi^2$ distribution on the ($f_0$,$d_0$) plane. The optimized  ($f_0$,$d_0$), marked  by red cross, converge to the input one (yellow star). The consistency between the input values and the reconstructed ones in panel (a), (b) and (d) demonstrates the validity of the deblurring algorithm.

\begin{figure}[hptb]
    \centering
    \includegraphics[width=0.42 \textwidth]{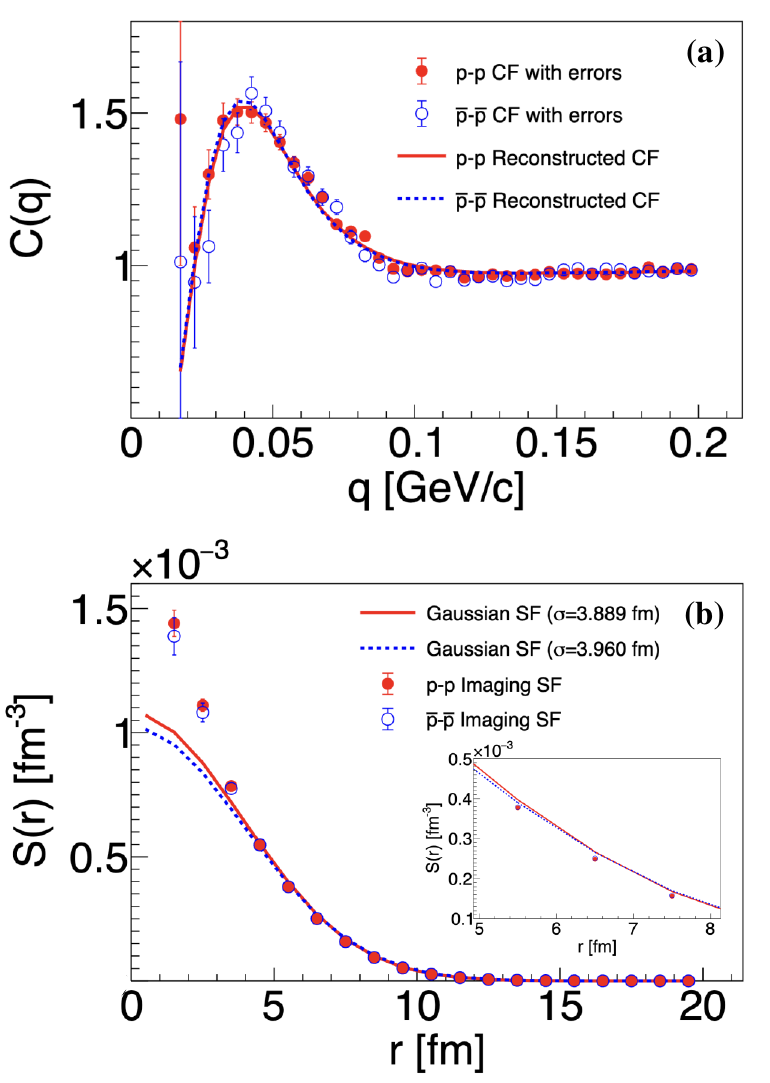}
    \caption{(Color Online) (a) Reconstructed correlation functions (CFs) for $pp$ (red solid curve) and $\bar{p}\bar{p}$ (blue dashed curve) pairs, compared with experimental data from \cite{Adamczyk2015} (red dots for pp and blue circles for $\bar{p}\bar{p}$ pairs).
    (b) Reconstructed SFs for $p$ (red dots) and $\bar{p}$  (blue circles) in Au+Au at center of mass energy 200 GeV. The symbols are the results of direct imaging without Gaussian assumption. The SFs with Gaussian shape obtained in \cite{Adamczyk2015} are shown by the curves for reference.}
    \label{FixedImagingCompare}
\end{figure}

{\it Application to data -} Now we apply the method to the $pp$ and $\bar{p}\bar{p}$ CFs in Au+Au at  $\sqrt{S_{\rm NN}}=200 ~\rm{GeV}$  \cite{Zhang2017}, taken by STAR collaboration \cite{Adamczyk2015,Ackermann2003} at  the Relativistic Heavy Ion Collider (RHIC) \cite{Harrison2003,STAR2016}. The experimental CFs are inclusive and contain the contributions from the residual CFs of $p\Lambda$ and $\Lambda\Lambda$. The residual contribution, which shall be subtracted,  is derived by the difference between the two fitting curves  with and without the feedback corrections shown in  \cite{Zhang2017}. The pure $pp$ and $\bar{p}\bar{p}$ CFs  are presented in Fig. \ref{FixedImagingCompare} (a) by the symbols, where the uncertainties are taken same as in the experiment for each data point.

In the calculation of  the Kernel $K_{ij}$, one needs the momentum distribution of $p$ and $\bar{p}$. The transverse momentum $p_{\rm t}$ distributions are taken from the midrapidity range in the same colliding system \cite{PHENIX2022} measured by PHENIX experiment \cite{Adler2004,Adcox2003}.  The longitudinal rapidity $y_{\rm l}$ as well as the azimuth in transverse direction are uniformly distributed. The cuts  $0.4<p_{\rm t}< 2.5 \rm {GeV}/c$ and  $|y_{\rm l}|<0.7$ are taken as same as STAR data where the CFs are measured. As  well noticed, the LL model provides relatively weak constraints on  $d_0$  when analyzing experimental data. And the variation of $d_0$ brings correspondingly uncertainty to the source function to be extracted.  So the widely accepted theoretical and experimental approximation $d_0 = 2.8\rm ~fm$ is fixed here for both  $pp$  and  $\bar{p}\bar{p}$  pairs. The convergence criterion is set to $1\times10^{-2}$. The corresponding optimal values of  $f_0$  obtained are as follows:
\begin{equation}
    f_{0(pp)}=(7.7\pm 0.2) \text{fm},\quad f_{0(\bar{p}\bar{p})}=(7.9\pm 0.3) \text{fm}.
\end{equation}
 The  $f_0$  values for  $pp$  and  $\bar{p}\bar{p}$  pairs are identical and consistent with both the theoretical calculations and the experimental results from STAR collaboration  \cite{Adamczyk2015}. 

The converged CFs of $pp$  and  $\bar{p}\bar{p}$ reproduce fairly the experimental data, as demonstrated by the solid and dashed curves in  Fig. \ref{FixedImagingCompare} (a), respectively. The results of the restored source imaging are plotted in panel (b) by red dots for $p$  and blue circles for $\bar{p}$. In order to estimate the uncertainty of each imaging point, numerous pseudo-CFs for $pp$ and $\bar{p}\bar{p}$ are sampled, smearing the content in each individual $q$ bin by Gaussian fluctuation, for which the center (standard deviation) is taken from the experimental value (uncertainty). Then the standard deviation of the reconstructed SFs are calculated and plotted on each data point bin-by-bin as the error bar on the SF. 

The direct imaging result in Fig.~\ref{FixedImagingCompare} (b) has important implications.  First, Within the uncertainty, protons and antiprotons share the identical source distribution at the instant of freeze-out. In addition to the momentum distribution analysis conducted in Ref. \cite{Zhang2017}, the direct imaging results presented here provide the final piece of evidence supporting matter-antimatter symmetry in coordinate space, as it pertains to their production and evolution in high-energy HICs. Notably, this evidence is obtained without imposing any assumptions regarding the shape of the source. Second, the profile of the reconstructed SFs deviates from the Gaussian distribution for both protons and antiprotons.   Quantitative calculation yields the  probability for the SF to be Gaussian is  $0$  and  $2.2 \times 10^{-14}$ for  $p$  and  $\bar{p}$, respectively. For a clear comparison,  the Gaussian distribution obtained in \cite{Zhang2017} are  shown by the curves in panel (b), respectively. The emission source is more concentrated at smaller radii ($r<5$ fm) while showing reduced density at larger radii ($5<r<8$ fm), as highlighted in the inset. It is consistent with the picture of an extremely fast collision. Namely,  protons and antiprotons, formed by the coalescence of quarks and antiquarks created inside the overlapping region, are not fully randomized in coordinate space prior to the freeze-out instant, otherwise they are expected to extend to the outer space in Gaussian distribution as a deduction of the central limit theorem. 
Our work offers a support to the recent idea of probing the initial shape and fluctuation of the nucleus via relativistic HIC
\cite{Zhao2024,Jia2023,Zhang2022,Jia2022,Giacalone2021,Liu2024,Jia2024,Schenke2024,Wang2024b}, since the collision is so rapid that the spatial information of the colliding nuclei is partially preserved.

The impact of the bin-by-bin variation of the SF on the CF has been analyzed. It is observed that variations in the first few bins of the source function (except for the first bin which is below the low limit cut in LL model calculation) significantly impact the CF, whereas the variations at larger  $r$  values are less influential. Inversely, the experimental uncertainty of the CFs near the peak ($q\approx 40 ~\rm MeV$) brings considerable variation to the SF. In order to achieve more accurate imaging, it is demanded to further reduce the uncertainty of the CF, particularly at small $q$.

The method developed in this work holds significant potential for important applications. Given high-precision experimental CFs, this technique provides a novel approach to identify exotic nuclear structures, particularly in future radioactive ion beam facilities. For instance, it could be employed to probe the proton density profile of bubble nuclei—a recently proposed phenomenon \cite{Mutschler2017, Natowitz2024} that remains under debate\cite{Chen2024}. Furthermore, when extended to three dimensions, this method could be applied at facilities such as RHIC to investigate higher-order deformations of atomic nuclei, which are typically studied through nuclear spectroscopy \cite{Frauendorf2024,Wei2024}.

{\it Summary - } A direct imaging method  on femtometer scale has been realized by combining the Richardson-Lucy deblurring algorithm and the intensity interferometry in  relativistic heavy ion collisions. This method allows for a more accurate extraction of source distributions and final state  interaction  between particle pair based on the  Lednick\'y-Lyuboshitz model analysis. Applying this imaging method to $pp$  and  $\bar{p}\bar{p}$ correlation functions in Au+Au at $\sqrt{S_{\rm NN}}=200$  GeV measured by the STAR collaboration,  we observe 
the same freeze-out source functions  for  $p$  and  $\bar{p}$ emissions with distinct non-Gaussian characteristics. These findings confirm the matter-antimatter symmetry in coordinate space once produced and suggest the ultrafast feature of the evolution of the collision, which is hardly accessible through other experimental means alone. This method can be extended to three-dimensional imaging for further applications and is inherently compatible with machine-learning approaches \cite{Wang2024c}, opening new avenues for high-precision source reconstruction in heavy-ion collisions. It represents a significant advancement of our ability to image the femtoscopic heavy ion collisions and to explore matter-antimatter physics in terrestrial laboratories.

{\it  Acknowledgements~}
This work is supported by the National Natural Science Foundation of China under Grant Nos.12335008 and 12205160. The authors acknowledge the support of the Center for High Performance Computing and Initiative Scientific Research Program in Tsinghua University. Sincere gratitude is extended to Nu Xu and Pengfei Zhuang for their valuable suggestions and insightful discussions. 

{\it Author contributions~}
JHX, code developing and calculation. ZQ, code reviewing. YJW, analysis and result checking. ZGX, project leading and supervising. All authors contribute to discussions, draft writing and editing.


\end{document}